\documentclass[reprint,noshowpacs,noshowkeys,prd,balancelastpage,nofootinbib]{revtex4}
\usepackage{amsfonts}
\usepackage{mdframed}
\usepackage{amssymb}
\usepackage{footnote}
\usepackage{amsmath}
\usepackage{graphicx}
\usepackage{float}
\usepackage[font={footnotesize,it}]{caption}
\usepackage[utf8]{inputenc}
\usepackage{natbib}
\usepackage{tikz}
\usepackage{tikz-3dplot}
\usepackage[colorlinks=true,
            linkcolor=purple,
            urlcolor=purple,
            citecolor=blue]{hyperref}

\begin{document}

\title{One-dimensional quantum scattering from multiple Dirac $\delta -$%
potentials: A Python-based solution }
\author{Erfan Keshavarz }
\email{erfan.keshavarz@emu.edu.tr}
\author{S. Habib Mazharimousavi}
\email{habib.mazhari@emu.edu.tr}
\affiliation{Department of Physics, Faculty of Arts and Sciences, Eastern Mediterranean
University, Famagusta, North Cyprus via Mersin 10, T\"{u}rkiye}
\date{\today }

\begin{abstract}
In this research, we present a Python-based solution designed to simulate a
one-dimensional quantum system that incorporates multiple Dirac $\delta -$%
potentials. The primary aim of this research is to investigate the
scattering problem within such a system. By developing this program, we can
generate wave functions throughout the system and compute transmission and
reflection amplitudes analytically and numerically for an infinite range of
combinations involving potential strengths, potential separations, and the
number of potential centers in the form of the Dirac $\delta -$functions.
Furthermore, by modifying the code, we investigate the so-called
"transmission resonances" which yields the energy of the quantum particles
undergoing a perfect transmission. Subsequently, our research is extended by
considering impurities in the system.
\end{abstract}

\maketitle

\section{Introduction}

The Dirac $\delta -$potential profoundly impacts the field of science, with
significant applications in various areas. For instance, the Kronig-Penny
model stands out as a crucial example, as it effectively elucidates the
formation of band gaps in crystal structures \cite{E1}. The delta potential,
often portrayed using the Dirac delta function, finds significant utility in
quantum mechanics when describing interactions within systems of weakly
interacting bosons. A prominent example of this application can be seen in
the study of cold atomic gases, particularly in the context of Bose-Einstein
condensates \cite{E2}. In this context, the delta potential is a
mathematical tool to model the localized potential energy at a specific
point or region in space, which characterizes the interaction between the
weakly interacting bosons \cite{E3}.

Indeed, the concept of delta potentials and their successful applications
have led to significant research in various areas of physics (see for
instance \cite{D1,D2,D3,D4,D5,D6,D7,D8,D9}). One noteworthy study in this
domain demonstrates that scattering and reflection amplitudes of an
arbitrary potential can be approximated using delta potentials \cite{E4,E5}.
An important feature of Dirac $\delta -$potential is their exact
solvability, which renders them highly suitable for educational purposes 
\cite{E6}. Reference \cite{E7} offers insights into Green's functions and
the solution of the Lippmann-Schwinger equation for a single Dirac $\delta -$%
potential. Furthermore, reference \cite{E8} takes a pedagogical approach to
explore multiple scattering theory for double delta centers using the
Lippmann-Schwinger equation. These educational materials contribute to a
deeper understanding of fundamental quantum mechanics and scattering theory
concepts. A recent review \cite{E4} has shed light on some fascinating
characteristics of one-dimensional Dirac $\delta -$potential. It
specifically explores the spectrum of continuum and bound states within
delta potentials and other potentials amenable to exact solutions.

Moreover, the review delves into the study of multiple $\delta $-function
potentials in Fourier space and frames the bound state problem in terms of a
matrix eigenvalue problem. In recent years, there has been a growing
interest in the study of one-dimensional systems featuring multiple Dirac $%
\delta -$potential. Researchers have applied transfer matrix techniques to
explore fascinating scattering phenomena, including transmission resonances
(occurring at energies with a transmission amplitude of one), threshold
anomalies (where the reflection amplitude approaches zero under certain
parameter conditions as the incoming particle's energy approaches zero), and
the investigation of Bloch states. These investigations have significantly
advanced our comprehension of wave behavior and particle interactions within
such systems \cite{E9,E10,E11,E12,E13,E14}. Pereyra et al \cite{E15,E16}
used a transfer matrix to formulate a theory for the finite periodic system.
In addition, there has been a study on the use of transfer matrices for Schr%
\"{o}dinger electrons in ordered and disordered systems \cite{E17}.
Moreover, for Schr\"{o}dinger electrons, analytic solutions based on full
matrix transfer were studied in \cite{E18}. Additionally, the case of 1D
Dirac-like problems for bound states and total transmission based on the
transfer matrix was studied in \cite{E19}. Furthermore, Coquelin et al. \cite%
{E20,E21} carried out experiments on electron transmission through a finite
biperiodic GaAs/AlGaAs superlattice consisting of alternating types of unit
cells. Moreover, transmission through biperiodic semiconductor superlattices
was studied based on the transfer matrix method \cite{E22}.

In this study, we initiate our exploration by representing the Schr\"{o}%
dinger equation in a dimensionless form. Furthermore, we revisit the topic
of scattering, specifically focusing on both single and double Dirac $\delta
-$potential in Sec. II. Subsequently, our study advances to the simulation
of a system consisting of multiple one-dimensional Dirac $\delta -$
potentials using Python in Sec. III. This program exhibits remarkable
versatility, accommodating any desired number of potentials. Our focus then
shifts towards enhancing the program's capabilities by modifying the code to
generate regional wavefunctions. By imposing appropriate boundary conditions
at each potential point, we establish a system of equations that allows us
to determine transmission and reflection amplitudes. This fundamental
analysis forms the basis for understanding the behavior of quantum particles
in our system. To further enrich our investigation, we extend our code to
explore transmission resonances within the system. This extension enables us
to obtain the exact energy of the quantum particle at which total
transmission occurs, shedding light on critical aspects of the system's
behavior. Additionally, by generalizing the code, we delve into the study of
impurities within the system. This generalized approach enables us to
investigate how impurities impact the behavior of quantum particles in the
presence of multiple Dirac $\delta -$potential. This exploration deepens our
understanding of real-world scenarios where imperfections and variations
exist. We bring together the comprehensive insights gained from our
simulations and analyses to provide a general analytical solution for
transmission and reflection probabilities in the context of scattering from
multiple Dirac $\delta -$potential in Sec. IV. We conclude our research in
Sec. V.

\section{Revisiting scattering from single and double Dirac $\protect\delta %
- $potential}

\begin{figure}[tbph]
\includegraphics[width=85mm,scale=1]{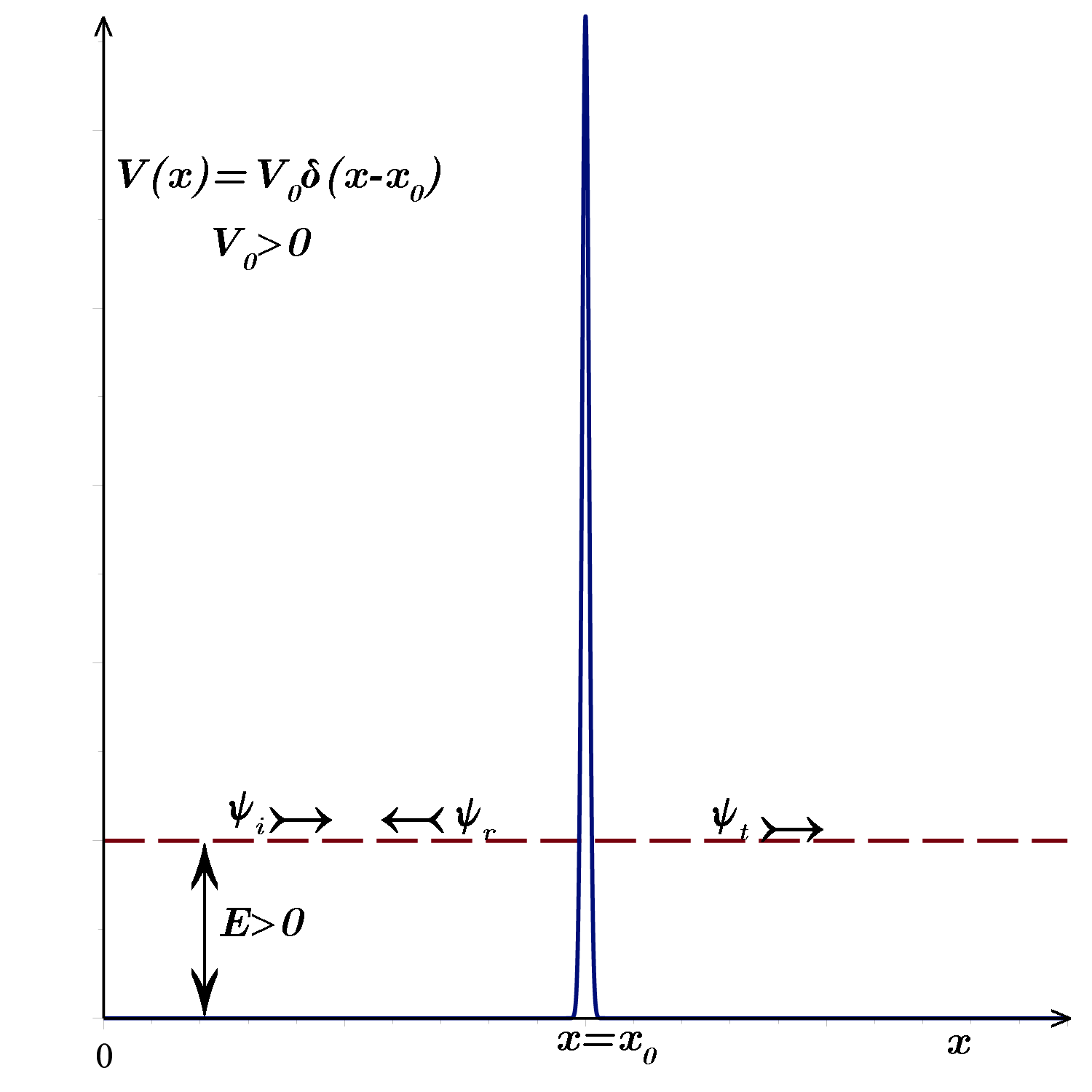} %
\includegraphics[width=85mm,scale=1]{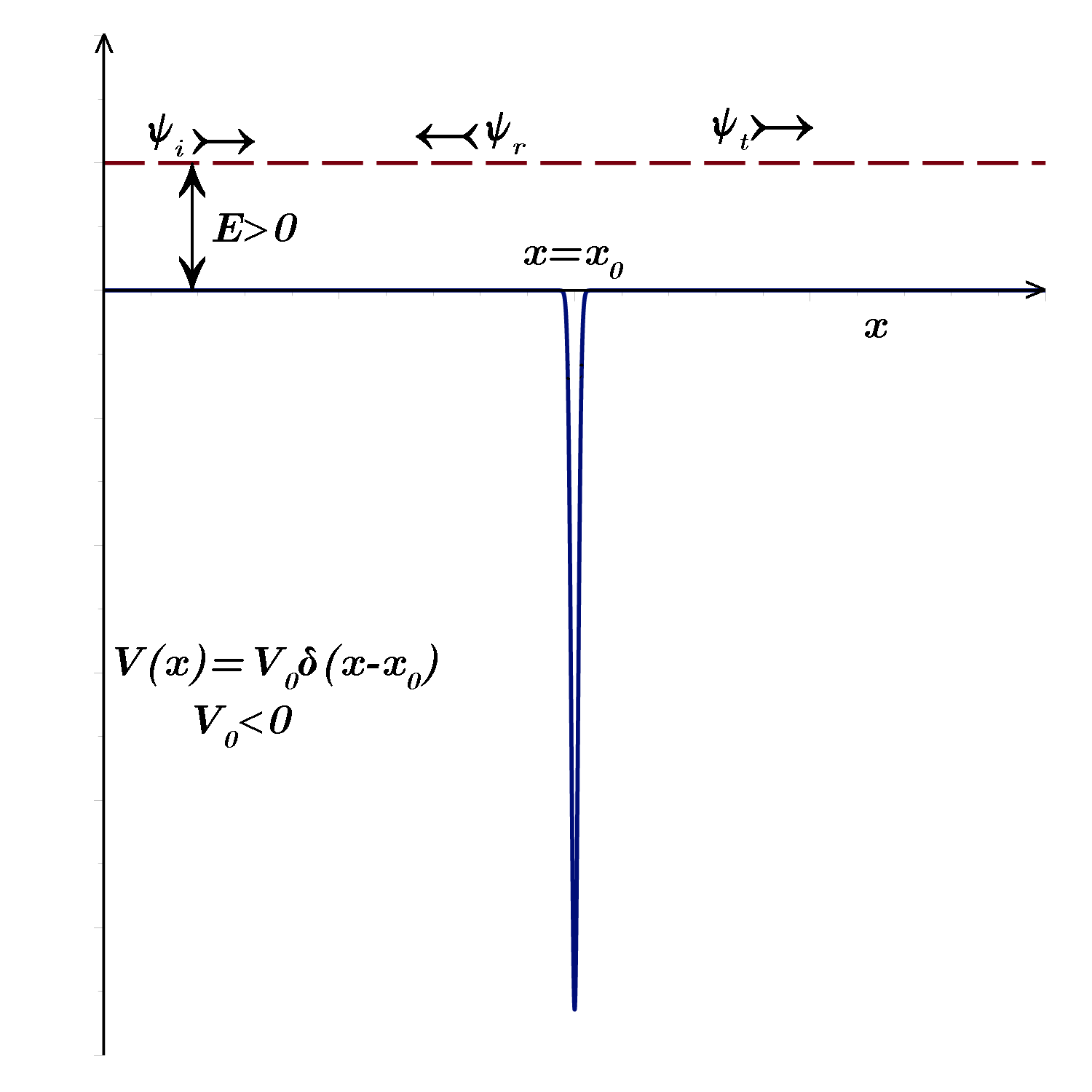}
\caption{A generic scheme of scattering from a single Dirac $\protect\delta %
- $potential. The sign of the potential strength $V_{0}$ determines whether
the system consists of a well or a barrier. The dashed line represents the
energy of the quantum particle which is positive and continuous for the
scattering in the configuration of this work.}
\label{F1}
\end{figure}

\textbf{I}n this section, we consider the one-dimensional Schr\"{o}dinger
equation with a Dirac $\delta -$potential expressed by 
\begin{equation}
V\left( x\right) =V_{0}\delta \left( x-x_{0}\right) ,  \label{R1}
\end{equation}%
in which $V_{0}$\ and $x_{0}$\ are the strength and the center of the
potential, respectively, as is depicted in Fig. \ref{F1}. Subsequently, we
discuss the scattering of a quantum particle with a positive energy $E$, to
determine the transmission and reflection amplitudes associated with the
system. Let us start with the time-independent Schr\"{o}dinger equation 
\begin{equation}
-\frac{\hbar ^{2}}{2m}\frac{d^{2}\psi }{dx^{2}}+V_{0}\delta \left(
x-x_{0}\right) \psi =E\psi ,  \label{1}
\end{equation}%
which describes the wave function of a quantum particle of mass $m$\ and
energy $E>0$\ undergoing the scattering potential (\ref{R1}). Furthermore,
we make the Schr\"{o}dinger equation dimensionless by introducing a new
variable, 
\begin{equation}
y=kx,  \label{R2}
\end{equation}%
in which $k^{2}=\frac{2mE}{\hbar }$. Hence, (\ref{1}) simplifies as 
\begin{equation}
-\frac{d^{2}\psi }{dy^{2}}+\xi \delta (y-y_{0})\psi =\psi ,  \label{2}
\end{equation}%
where $\xi =\frac{\tilde{V_{0}}}{k}$\ is a dimensionless parameter and $%
\tilde{V_{0}}=\frac{2mV_{0}}{\hbar ^{2}}$. We add that the dimension of $%
V_{0}$\ in (\ref{R1}) is not "energy" and instead it is "energy$\times $%
length". By integrating both sides of (\ref{2}) from $y_{0}-\epsilon $ to $%
y_{0}+\epsilon $ and calculating its limit as $\epsilon $ approaches zero,
one finds 
\begin{equation}
\left. \frac{d\psi {}_{R}}{dy}\right\vert _{y=y_{0}}{\small -}\left. {\small 
\frac{d\psi {}_{L}}{dy}}\right\vert _{y=y_{0}}{\small =\xi }\left. {\small %
\psi }\right\vert _{y=y_{0}}{\small .}  \label{3}
\end{equation}

\begin{figure}[tbph]
\includegraphics[width=100mm,scale=1]{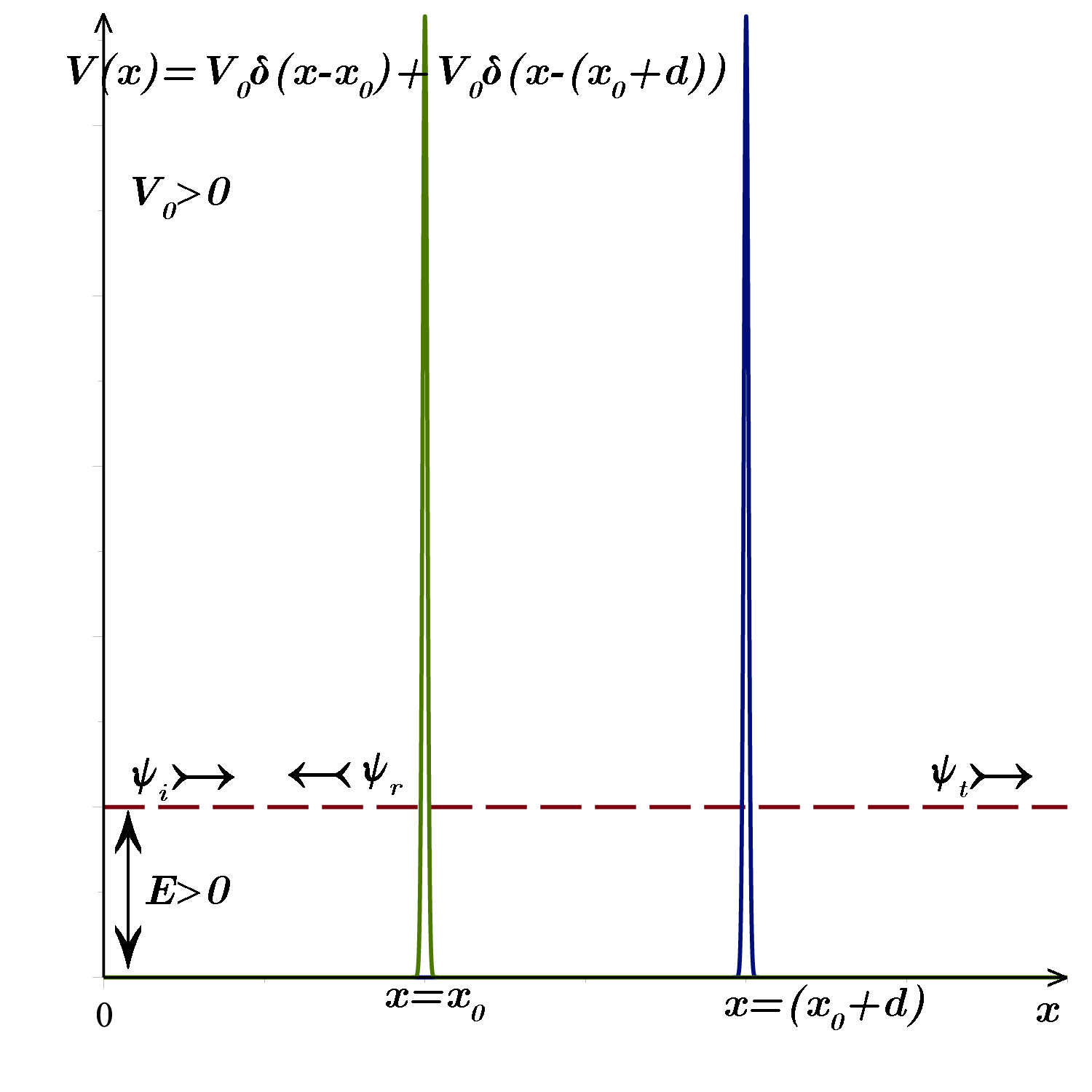}
\caption{A generic scheme of scattering from a double Dirac $\protect\delta %
- $potential separated by a distance $d$. The dashed line represents the
positive and continuous energy of the quantum particle.}
\label{F2}
\end{figure}

The latter equation depicts the discontinuity of the first derivative of the
wave function at the center of the Dirac $\delta -$potential. We also recall
that the wave function is continuous everywhere i.e., $\left( \psi
{}_{R}=\psi {}_{L}\right) _{y=y_{0}}$. Herein $\left. \frac{d\psi {}_{R}}{dy}%
\right\vert _{y=y_{0}}$ and $\left. {\small \frac{d\psi {}_{L}}{dy}}%
\right\vert _{y=y_{0}}$ represent the derivative of the wave function from
the right-hand side and the left-hand side of $y=y_{0}$, respectively. Now
by solving the Schr\"{o}dinger equation in regions $y<y_{0}$ (Left-side) and 
$y_{0}<y$ (Right-side), we obtain 
\begin{equation}
\psi _{L}=\exp (iy)+r\exp (-iy),  \label{4}
\end{equation}%
and 
\begin{equation}
\psi _{R}=t\exp (iy),  \label{5}
\end{equation}%
in which $r$ and $t$ are the so-called reflection and transmission
amplitudes. One notes that the wave functions of the free particle in the
left and the right of the Dirac $\delta -$potential are not normalizable. By
applying the continuity condition of the wave function and the discontinuity
condition of the first derivative of the wave function i.e., (\ref{3}) at $%
y=y_{0}$, we determine the transmission and reflection amplitudes which are
expressed by 
\begin{equation}
t=\frac{2i}{2i-\xi },  \label{6}
\end{equation}%
and 
\begin{equation}
r=\frac{\xi \exp (2iy_{0})}{2i-\xi }.  \label{7}
\end{equation}%
The conservation of the particle implies that $\left\vert t\right\vert
^{2}+\left\vert r\right\vert ^{2}=1$ in which $\left\vert t\right\vert ^{2}$
and $\left\vert r\right\vert ^{2}$ are the transmission and the reflection
probabilities of the quantum particle scattered from the single Dirac $%
\delta -$potential (\ref{R1}). In Fig. \ref{F2} a double Dirac $\delta -$%
potential is depicted such that the first $\delta -$potential is located at $%
x_{0}=0$\ and the second one is separated by a distance $d$\ from the first
and both potentials are equal in strength i.e., $V_{01}=V_{02}=V_{0\text{ }}$%
\ and consequently $\xi _{1}=\xi _{2}=\xi =\frac{\tilde{V}_{0}}{k}$. By
executing a similar procedures for the double Dirac $\delta -$potential as
illustrated in Fig. \ref{F2}, we derive the transmission and reflection
amplitudes that are respectively given by 
\begin{equation}
t=\frac{4}{\xi ^{2}(\exp (2i\tilde{d})-1)+4(i\xi +1)},  \label{8}
\end{equation}%
and 
\begin{equation}
r=\frac{\xi ^{2}(1-\exp (2i\tilde{d}))-2i\xi (\exp (2i\tilde{d})+1)}{\xi
^{2}(\exp (2i\tilde{d})-1)+4(i\xi +1)},  \label{9}
\end{equation}%
in which $\tilde{d}=kd$\ is the dimensionless separation parameter.{} 

\section{Scattering from multiple Dirac $\protect\delta -$potentials: Python
solution}

The issue arises when attempting to generalize the system to encompass
multiple Dirac $\delta -$ potentials. Analyzing scattering in such a system
can become tedious and time-consuming. Solving the Schr\"{o}dinger equation
throughout the system entails the application of continuity and
discontinuity boundary conditions at each potential site. This process
necessitates solving a system of equations to determine the transmission and
reflection amplitudes. In this section, we propose a Python program to
explore the one-dimensional scattering problem involving multiple Dirac $%
\delta -$potential, categorized into various scenarios. These categories
include examining whether the system has equidistantly spaced potentials or
not, whether the potentials are of equal strength, or if they contain
impurities. The project's code is hosted on GitHub, where a repository
containing comprehensive documentation and a detailed report outlines the
code's functionality and usage. For further exploration, visit \cite{GitHub}.

\subsection{Wave functions and IVP}

The program contains several libraries such as NumPy, SymPy, SciPy for
numerical and symbolic computation, and finally, Matplotlib for graphical
visualization. To initiate the system of equations, the first step involves
establishing the initial values, namely the potential\_list and
distance\_list. The program itself is implemented as a user input program,
offering various options to the user. Initially, the user is presented with
a choice between equal distances or non-equal distances, as well as equal
potentials or non-equal potentials. Furthermore, the user is prompted to
input the value of $k$. With these inputs at hand, we proceed to compute the
list of $\xi _{i}$. These options allow for flexibility in defining the
characteristics of the generated multiple Dirac $\delta -$potential. Once
you have selected your preferred option and set the initial values, such as $%
\xi _{i}$ and distances $d_{i}$, the code will initiate by providing you
with a list of general wave functions and their corresponding derivatives as
functions of $y$.

\subsection{Boundary conditions, transmission and reflection amplitudes}

Once the wave functions and their corresponding derivatives have been
compiled into a list, it is crucial to independently apply the boundary
conditions for each distinct region, considering both equal and non-equal
distances. After we have algebraically defined the boundary conditions and
have obtained the initial values of $\xi _{i}$\ and distances $d_{i}$, the
next step is to import these values and construct a system of equations.
Once the system of equations is formed, we can proceed to solve it
effectively. Once the solutions for each amplitude in every region have been
obtained, the next step is to achieve the transmission and reflection
probabilities, and finally, one can check the conservation of particles.
Additionally, one may assign amplitudes to the wave functions to calculate
their conjugates as well as the absolute value square of the wave functions.

As an example, let us consider a system containing double Dirac $\delta -$%
potentials with $n=2$, $\tilde{V}_{01}=\tilde{V}_{02}=$ $\tilde{V_{0}}=1$, $%
k=1$, and $d=1$ (and therefore $\xi =$ $\tilde{d}=1$). By executing the
code, we obtain the following wave functions and their corresponding
derivatives within the regions 1, 2, and 3 (see Fig. \ref{F2}). This outcome
reflects the results of the implemented procedures, allowing for a clear
examination of the wave functions and their derivatives in these specific
regions as stated below%
\begin{equation}
\psi \left( y\right) =\left\{ 
\begin{array}{lll}
\exp (iy)+r\exp (-iy), &  & \text{Region 1} \\ 
a_{2}\exp (iy)+b_{2}\exp (-iy), &  & \text{Region 2} \\ 
t\exp (iy), &  & \text{Region 3}%
\end{array}%
\right. ,  \label{10}
\end{equation}%
and%
\begin{equation}
\frac{d\psi \left( y\right) }{dy}=\left\{ 
\begin{array}{lll}
i\exp (iy)-ir\exp (-iy), &  & \text{Region 1} \\ 
ia_{2}\exp (iy)-ib_{2}\exp (-iy), &  & \text{Region 2} \\ 
it\exp (iy), &  & \text{Region 3}%
\end{array}%
\right. ,  \label{11}
\end{equation}%
in which $a_{2},$\ and $b_{2}$\ are some integration constants. The provided
code applies the boundary conditions at the potential sites, inserts the
initial values, and constructs a system of equations as follows 
\begin{equation}
\left\{ 
\begin{array}{l}
a_{2}+ia_{2}+b_{2}-ib_{2}+ir-i=0 \\ 
-ia_{2}\exp (i)+ib_{2}\exp (-i)+it\exp (i)+t\exp (i)=0 \\ 
ia_{2}-ib_{2}-\xi (a_{2}+b_{2})+ir-i=0 \\ 
-ia_{2}\exp (i)+ib_{2}\exp (-i)-\xi t\exp (i)+it\exp (i)=0%
\end{array}%
\right. .  \label{16}
\end{equation}%
Moreover, the code solves the system of equations and determines the unknown
parameters i.e. $a_{2}$\ and $b_{2}$\ as well as the transmission and
reflection amplitudes $t$\ and $r$. These calculations are essential for a
comprehensive understanding of the system's behavior and how it responds to
the given configuration and potential sites. For the particular example in
this section the code yields%
\begin{equation}
\left\{ 
\begin{array}{l}
r=-0.0597-0.690i \\ 
t=0.336-0.638i \\ 
a_{2}=0.655-0.470i \\ 
b_{2}=0.2854-0.220i%
\end{array}%
\right. ,  \label{17}
\end{equation}%
in which for the sake of readability, the numerical quantities are reported
up to 3-decimal. For further accuracy, one may use the program. It is easy
to verify that the conservation of the particle is held i.e., $\left\vert
t\right\vert ^{2}+\left\vert r\right\vert ^{2}=1$.

\section{Graphical representation and transmission resonance}

In this section, we divide the code into distinct components, each
addressing different scenarios. Then, by solving analytically the system of
equations determined from the boundary conditions, the code obtains the
transmission and reflection amplitudes for any given number of potentials.
This enables the code to create graphical plots of transmission and
reflection probabilities by taking into account their distances and the
strengths. Additionally, it allows for determining the energy of the quantum
particle that the particle requires for total transmission i.e., $\left\vert
t\right\vert ^{2}=1$.

\subsection{Equal distances and equal potentials i.e., $d_{i}=d$ and $%
V_{0i}=V_{0}$}

Initially, we proceed by inserting the values of amplitudes into each
regional wave function. Then, we define a range for our scattering problem,
which includes potentials at user-defined distances. The code will then plot
the absolute value square of the wave function (see \cite{GitHub}) within
the scattering problem in hand. As we progress with the code, our initial
step involves solving the system of equations analytically to obtain the
transmission and reflection amplitudes, considering the parameters $d$ and $%
\xi $. Subsequently, we visualize the scattering behavior of particles by
plotting the transmission and reflection probabilities. 
\begin{figure}[tbph]
\includegraphics[width=100mm,scale=1]{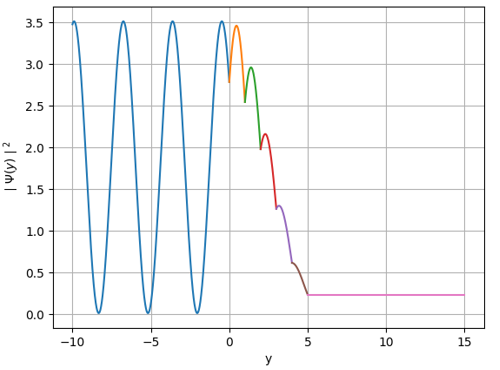}
\caption{Plot of $\left\vert \protect\psi \right\vert ^{2}$ in terms of $y$
for $\tilde{V}=1$, $k=1$, $d=1$, and $y_{0}=0$. As is seen in the figure, at
the location of the Dirac $\protect\delta -$potentials i.e., $y=0,1,2,3,4,$
and $5$ the derivative of the wave function is discontinuous.}
\label{F3}
\end{figure}

As an illustrative example, we execute the provided code for a system of six 
$\delta -$potentially with equal strength and distances. Firstly the
analytical solutions for transmission and reflection amplitudes are as
follows 
\begin{equation}
t=\frac{\alpha _{0}}{\sum_{i=0}^{6}\xi ^{i}\alpha _{i}},  \label{24}
\end{equation}%
\begin{equation}
r=\frac{\sum_{i=1}^{6}\xi ^{i}\beta _{i}}{\sum_{i=0}^{6}\xi ^{i}\alpha _{i}},
\label{25}
\end{equation}%
where%
\begin{equation}
\alpha _{i}=\left\{ 
\begin{array}{l}
64 \\ 
192i \\ 
80e^{2i\tilde{d}}+64e^{4i\tilde{d}}+48e^{6i\tilde{d}}+32e^{8i\tilde{d}%
}+16e^{10i\tilde{d}}-240 \\ 
160ie^{2i\tilde{d}}+64ie^{4i\tilde{d}}-32ie^{8i\tilde{d}}-32ie^{10i\tilde{d}%
}-160i \\ 
-120e^{2i\tilde{d}}+24e^{4i\tilde{d}}+48e^{6i\tilde{d}}+12e^{8i\tilde{d}%
}-24e^{10i\tilde{d}}+60 \\ 
-40ie^{2i\tilde{d}}+40ie^{4i\tilde{d}}-20ie^{8i\tilde{d}}+8ie^{10i\tilde{d}%
}+12i \\ 
5e^{2i\tilde{d}}-10e^{4i\tilde{d}}+10e^{6i\tilde{d}}-5e^{8i\tilde{d}}+e^{10i%
\tilde{d}}-1%
\end{array}%
\right. ,  \label{26}
\end{equation}%
and%
\begin{equation}
\beta _{i}=\left\{ 
\begin{array}{l}
-32ie^{2i\tilde{d}}-32ie^{4i\tilde{d}}-32ie^{6i\tilde{d}}-32ie^{8i\tilde{d}%
}-32ie^{10i\tilde{d}}-32i \\ 
48e^{2i\tilde{d}}+16e^{4i\tilde{d}}-16e^{6i\tilde{d}}-48e^{8i\tilde{d}%
}-80e^{10i\tilde{d}}+80 \\ 
-16ie^{2i\tilde{d}}-64ie^{4i\tilde{d}}-64ie^{6i\tilde{d}}-16ie^{8i\tilde{d}%
}+80ie^{10i\tilde{d}}+80i \\ 
56e^{2i\tilde{d}}+32e^{4i\tilde{d}}-32e^{6i\tilde{d}}-56e^{8i\tilde{d}%
}+40e^{10i\tilde{d}}-40 \\ 
30ie^{2i\tilde{d}}-20ie^{4i\tilde{d}}-20ie^{6i\tilde{d}}+30ie^{8i\tilde{d}%
}-10ie^{10i\tilde{d}}-10i \\ 
-5e^{2i\tilde{d}}+10e^{4i\tilde{d}}-10e^{6i\tilde{d}}+5e^{8i\tilde{d}}-e^{10i%
\tilde{d}}+1%
\end{array}%
\right. ,  \label{27}
\end{equation}%
with $\xi ^{i}=\frac{\tilde{V_{0i}}}{k}.$ In the sequel within some specific
numerical examples, we demonstrate how the code operates in different
configurations.

\subsubsection{$n=6,$ $\tilde{V}_{0}=1$ and $k=1$}

Now, by initializing the parameters as $\tilde{V}_{0}=1$, $k=1$, and $%
y_{0}=0 $, the code analyzes and visualizes the scattering problem, as
demonstrated in Fig. \ref{F3}. We note that in this figure $d=1$\ and the
graph displays the probability density i.e., $\left\vert \psi \right\vert
^{2}$\ in terms of $y.$\ The horizontal line to the right of the graph
implies the transmission probability i.e., $\left\vert t\right\vert ^{2}$\
which is a constant for the given configuration. Furthermore, as we have
mentioned the wave functions are not normalizable, and therefore the total
area under the curve $\left\vert \psi \right\vert ^{2}$\ is not supposed to
be one. The graph of $\left\vert \psi \right\vert ^{2}$\ in terms of $y$\
displays a relative probability distribution in different regions and also
the discontinuity of the derivative of the wave function and the location of
the Dirac $\delta -$potentials. The numerical values for the transmission
and reflection probabilities for this particular case are given by 
\begin{equation}
\left\vert t\right\vert ^{2}=0.236,  \label{39}
\end{equation}%
and 
\begin{equation}
\left\vert r\right\vert ^{2}=0.764.  \label{40}
\end{equation}

\begin{figure}[tbph]
\includegraphics[width=100mm,scale=1]{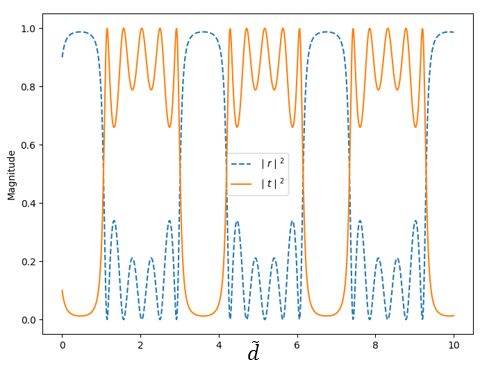}
\caption{This plot displays $\left\vert t\right\vert ^{2}$ and $\left\vert
r\right\vert ^{2}$ in terms of $\tilde{d}$ for $n=6$ and $\protect\xi =1.0$. 
}
\label{F4}
\end{figure}
\begin{figure}[tbph]
\includegraphics[width=100mm,scale=1]{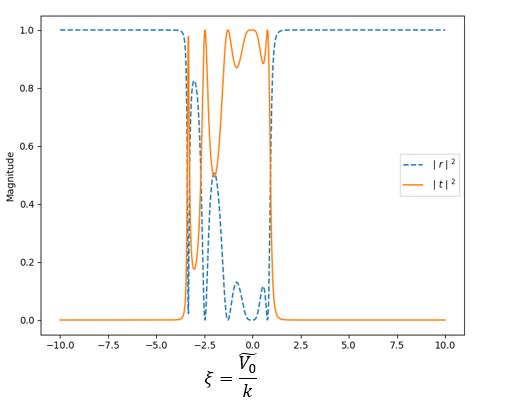}
\caption{This plot depicts $\left\vert t\right\vert ^{2}$ and $\left\vert
r\right\vert ^{2}$ in terms of $\protect\xi $ for $n=6$ and $\tilde{d}=1.0$.}
\label{F5}
\end{figure}
Furthermore, Fig. \ref{F4} and \ref{F5} illustrate the reflection and
transmission probabilities in terms of $\tilde{d}$\ and $\xi ,$\
respectively. From Fig. \ref{F4} the total transmission and reflection occur
multiple times at different energy levels and distances. Additionally, from %
\ref{F5}, we observe that as the ratio of potential to energy i.e., $\xi,$\
approaches to $2$\ and $-4$, the transmission probability tends to decrease
and approach zero. Total transmission and transmission resonance can occur
multiple times more frequently when $\xi $\ is negative (well).

To pinpoint the exact energy for the total transmission, we set $r=0$ and
then determine the value of $\xi $ with respect to the distances and energy.
For example, we explore the concept of total transmission in the context of 
\textit{double} Dirac $\delta -$potentials. In this case, the energy
associated with the total transmission may be expressed as

\begin{equation}
\xi =-\frac{2}{\tan (\tilde{d})}.  \label{41}
\end{equation}

\subsubsection{{} $n=6$, $\tilde{V}_{0}=1$, and $k=2$}

\begin{figure}[tbph]
\includegraphics[width=100mm,scale=1]{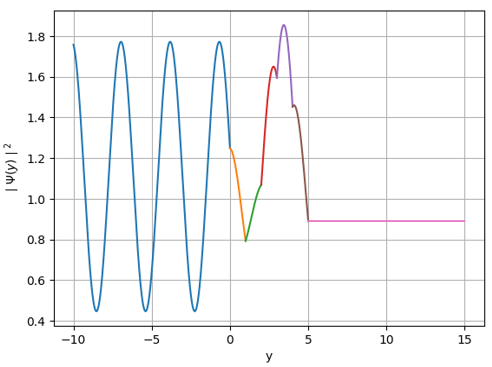}
\caption{Plots of $\left\vert \protect\psi \right\vert ^{2}$ in terms of $y$
for $n=6$, $\tilde{V}_{0}=1$, $d=1$ and $k=2$ }
\label{F6}
\end{figure}

\begin{figure}[tbph]
\includegraphics[width=100mm,scale=1]{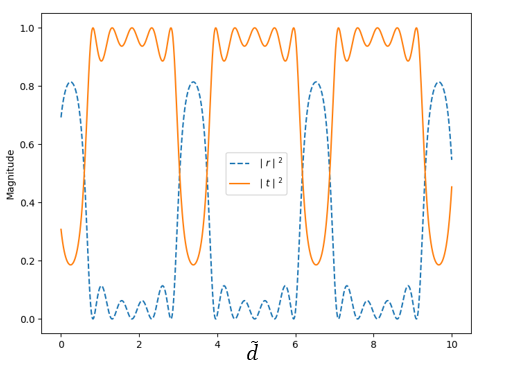}
\caption{Plots of the transmission and reflection probabilities in terms of $%
\tilde{d}$ for $\protect\xi =0.5.$}
\label{F7}
\end{figure}

As we increase the energy of the particle, we expect the transmission
probabilities to increase. It is observed from Figs. \ref{F6} and \ref{F7}
that as the energy of the particle increases, the probabilities of
transmission increase, while the probabilities of reflection decrease. In
the specific configuration with $n=6$, $\tilde{V}_{0}=1$, $d=1$ and $k=2$
the code yields%
\begin{equation*}
\left\vert t\right\vert ^{2}=0.8902,
\end{equation*}%
and 
\begin{equation}
\left\vert r\right\vert ^{2}=0.1097.  \label{47}
\end{equation}%
\bigskip {} 

\subsubsection{$n=6,$ $\tilde{V}_{0}=2$ and $k=1$}

\begin{figure}[tbph]
\includegraphics[width=100mm,scale=1]{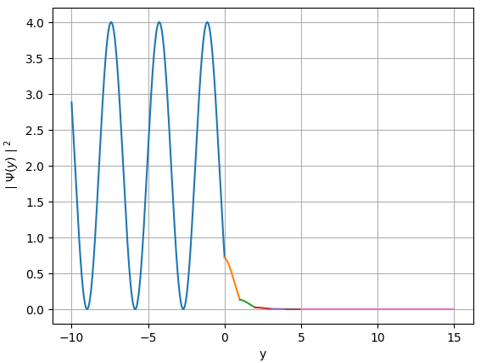}
\caption{Plot of $\left\vert \protect\psi \right\vert ^{2}$ in terms of $y$
for $\tilde{V}_{0}=2$, $d=1$ and $k=1$}
\label{F8}
\end{figure}

\begin{figure}[tbph]
\includegraphics[width=100mm,scale=1]{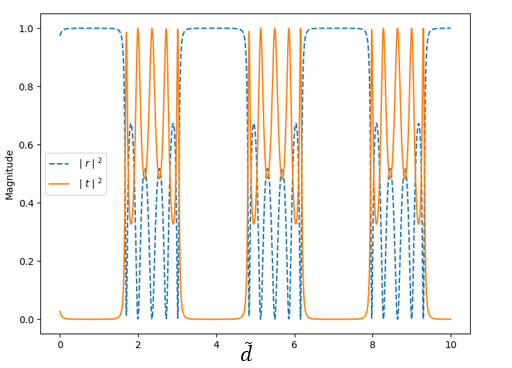}
\caption{Plot of the transmission and reflection probabilities in terms of $%
\tilde{d}$ for $\protect\xi =2.0.$}
\label{F9}
\end{figure}

From Figs. \ref{F8} and \ref{F9} as the strength of the potentials
increases, the probabilities of transmission decrease, while the
probabilities of reflection increase as illustrated below%
\begin{equation*}
\left\vert t\right\vert ^{2}=0.0001,
\end{equation*}%
and 
\begin{equation}
\left\vert r\right\vert ^{2}=0.9998.  \label{49}
\end{equation}


\subsection{Equal distances and non-equal potentials i.e., $d_{i}=d$}

Through code modification, one can introduce impurities in the potential
strengths, enabling the exploration of a system. By impurities, we mean that
out of several numbers of $\delta -$potentials with uniform strength, there
exist a few $\delta -$potentials with different strengths. By considering
these impurities, we have the opportunity to plot the absolute values square
of wave functions as well as the transmission and reflection probabilities.
This approach allows for a comprehensive analysis of the system's behavior
under the existence of impurities. The code modification is similar to the
previous cases. In the subsequent instances, we will examine specific
examples of these impurities, considering various scenarios to explore the
effects on the system.

\subsubsection{$n=8$, $\tilde{V_{01}}=0.1$, $\tilde{V_{02}}=...=\tilde{V_{08}%
}=1$, and $k=1$}

\begin{figure}[tbph]
\includegraphics[width=100mm,scale=1]{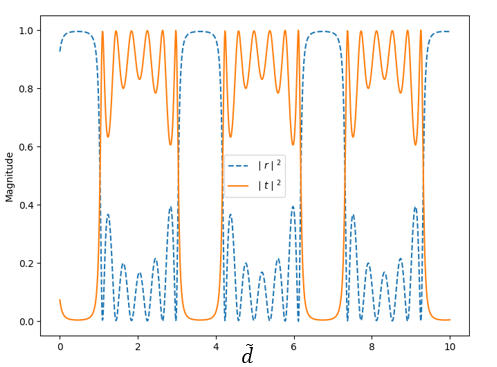}
\caption{In this figure we plot $\left\vert t\right\vert ^{2}$ and $%
\left\vert r\right\vert ^{2}$ in terms of $\tilde{d}$ for $n=8$, $\tilde{%
V_{01}}=0.1$, $\tilde{V_{02}}=...=\tilde{V_{08}}=1$, and $k=1$.}
\label{F10}
\end{figure}

When the impurities are considered comparatively weak, one may readily
disregard the influence of the impurities. For the case illustrated in Fig. %
\ref{F10}, $\tilde{V_{01}}=0.1$ is considered to be representing an impurity
in the system with its potential strength only 10\% of the rest. The
corresponding numerical values for transmission and reflection probabilities
with $d=1$ are as follows 
\begin{equation}
\left\vert t\right\vert ^{2}=0.284,
\end{equation}%
and%
\begin{equation}
\left\vert r\right\vert ^{2}=0.716.
\end{equation}

\subsubsection{$n=8,$ $\tilde{V}_{01}=0.5,\tilde{V}_{02}=...=\tilde{V}%
_{08}=1 $ and $k=1$}

\begin{figure}[tbph]
\includegraphics[width=100mm,scale=1]{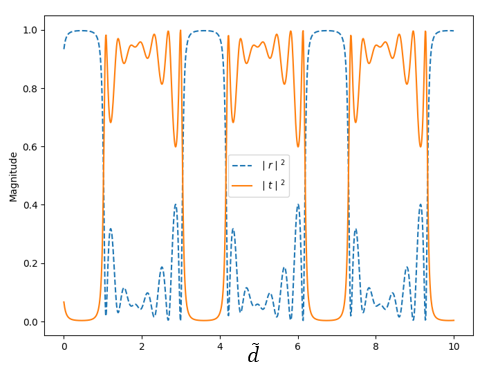}
\caption{Plots of the transmission and reflection probabilities in terms of $%
\tilde{d}$ for $n=8$, $\tilde{V_{1}}=0.5$, $\tilde{V_{2}}=...=\tilde{V_{8}}%
=1 $ and $k=1.$ }
\label{F11}
\end{figure}

By increasing the strength of the impurity potential in the system, the
probabilities of transmission and reflection exhibit significant differences
compared to previous case as shown in Fig. \ref{F11} (it should be compared
with Fig. \ref{F10}). The numerical values for transmission and reflection
probabilities with $d=1$ are given by%
\begin{equation}
\left\vert t\right\vert ^{2}=0.352,  \label{52}
\end{equation}%
and 
\begin{equation}
\left\vert r\right\vert ^{2}=0.6483.  \label{53}
\end{equation}

\begin{figure}[tbph]
\centering\includegraphics[width=85mm,scale=1]{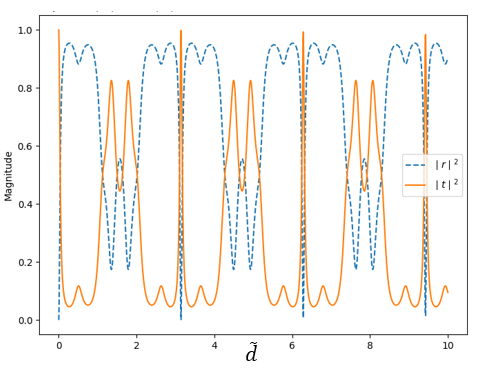}%
\includegraphics[width=85mm,scale=1]{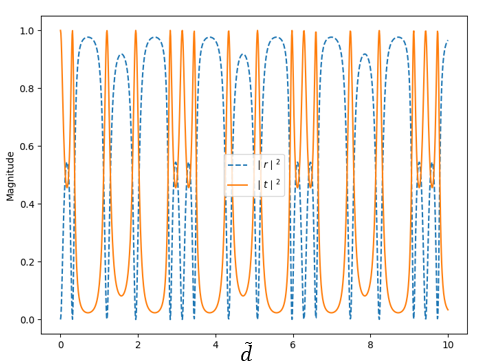} %
\includegraphics[width=85mm,scale=1]{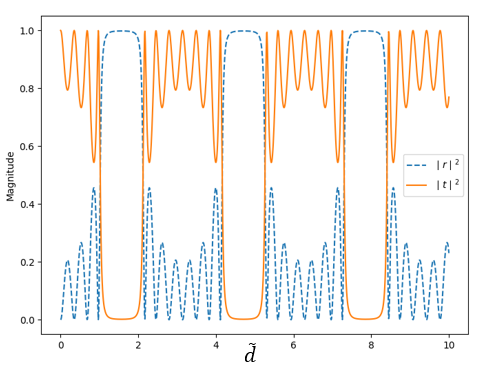}
\caption{The reflection and transmission operabilities in terms of $\tilde{d}
$ for $n=8$, $k=1$ and $\left( \tilde{V}_{0i}\right) =\left(
1,1,1,1,-1,-1,-1,-1\right) $ (upper-left panel), $\left( \tilde{V}%
_{0i}\right) =\left( 1,1,-1,-1,1,1,-1,-1\right) $ (upper-right panel) and $%
\left( \tilde{V}_{0i}\right) =\left( 1,-1,1,-1,1,-1,1,-1\right) $ (lower
panel).}
\label{F12}
\end{figure}

\bigskip Another interesting point to note is that in a system consisting of
multiple Dirac $\delta -$potential with different strengths, the order or
placement of these potentials causes changes in the probabilities of
transmission and reflection. For instance, in a system comprising of eight
potentials, with four being wells and the remaining four barriers, all with
a strength of $\tilde{V_{0i}}=1$, the placement of these potentials makes
significant impact on the results, as demonstrated in Fig. \ref{F12}. 

\subsection{Non-equal distances and non-equal potentials}

In this section, we extend the generality of our formalism by further
modifying the code. Let us consider scenarios where distances and potentials
are not constrained to be equal. By removing these constraints, we explore a
more diverse range of possibilities within the system. Despite these
variations, we are still able to calculate the transmission and reflection
amplitudes, allowing us to gain a comprehensive understanding of the
system's behavior under a more general configuration. Consider a system of
three potentials characterized by $\tilde{V_{01}},\tilde{V_{02}},\tilde{%
V_{03}},d_{1}$, $d_{2}$ and $k.$ Our analytic solutions for the transmission
and reflection amplitudes are given by 
\begin{equation}
t=\frac{-8i}{\gamma +\omega },
\end{equation}%
\begin{equation}
r=\frac{\lambda +\beta }{\gamma +\omega },
\end{equation}%
in which%
\begin{equation}
\left\{ 
\begin{array}{l}
\gamma =-\xi _{1}\xi _{2}\xi _{3}(e^{2i(\tilde{d}_{1}+\tilde{d}_{2})}-e^{2i%
\tilde{d}_{1}}-e^{2i\tilde{d}_{02}}+1)+2i\xi _{1}\xi _{2}(1-e^{2i\tilde{d}%
_{1}}) \\ 
\lambda =2i\xi _{2}\xi _{3}(e^{2i(\tilde{d}_{1}+\tilde{d}_{2})}-e^{2i\tilde{d%
}_{1}})-4(\xi _{1}+\xi _{2}e^{2i\tilde{d}_{1}}+\xi _{3}e^{2i(\tilde{d}_{1}+%
\tilde{d}_{2})}) \\ 
\beta =-\xi _{1}\xi _{2}\xi _{3}(e^{2i\tilde{d}_{1}}+e^{2i\tilde{d}%
_{2}}-e^{2i(\tilde{d}_{1}+\tilde{d}_{2})}-1)+2i\xi _{1}\xi _{2}(e^{2i\tilde{d%
}_{1}}-1)+2i\xi _{1}\xi _{3}(e^{2i(\tilde{d}_{1}+\tilde{d}_{2})}-1) \\ 
\omega =2i\xi _{1}\xi _{3}(1-e^{2i(\tilde{d}_{1}+\tilde{d}_{2})})+2i\xi
_{2}\xi _{3}(1-e^{2i\tilde{d}_{2}})+4(\xi _{1}+\xi _{2}+\xi _{3})-8i%
\end{array}%
\right. 
\end{equation}%
In a particular setting when $\tilde{V_{01}}=1,\tilde{V_{02}}=2,\tilde{V_{03}%
}=3$ and $d_{1}=1$, $d_{2}=2$ and $k=1,$ the transmission and reflection
amplitudes are obtained to be 
\begin{equation}
r=0.1434-0.908i,
\end{equation}%
and 
\begin{equation}
t=-0.391+0.025i.
\end{equation}%
{}

For another illustrative example, we consider a nonhomogeneous double Dirac $%
\delta -$potential. Within a detailed analysis, we generate analytical
expressions for the transmission and reflection amplitudes which are given
by 
\begin{equation}
t=\frac{4}{\xi _{1}\xi _{2}(e^{2i\tilde{d}}-1)+2i(\xi _{1}+\xi _{2})+4},
\end{equation}%
and 
\begin{equation}
r=\frac{\xi _{1}\xi _{2}(1-e^{2i\tilde{d}})-2i(\xi _{1}+\xi _{2}e^{2i\tilde{d%
}})}{\xi _{1}\xi _{2}(e^{2i\tilde{d}}-1)+2i(\xi _{1}+\xi _{2})+4},
\end{equation}%
in which\ $\xi _{i},k$ and $\tilde{d}$ are arbitrary. For a specific setting
of $\tilde{V_{01}}=1,\tilde{V_{02}}=-2,$ $k=2$ and $d=1$ we obtain%
\begin{equation}
r=-0.407+0.418i,
\end{equation}%
and%
\begin{equation}
t=0.775-0.240i.
\end{equation}%
The code allows users to input a wide range of variational values, enabling
the simulation of diverse scenarios with any desired number of potentials.

\section{Conclusion}

In this study, we embarked on an exploration of the one-dimensional form of
the multiple Dirac $\delta -$potentials. Utilizing Python programming, we
could design a quantum system consisting of numerous Dirac $\delta -$%
potential and investigate quantum scattering in various scenarios. The
program provided numerical and analytical solutions for the transmission and
reflection probabilities, accommodating any number of potentials. It also
simulated the wave function, providing a comprehensive view of the
scattering problem through graphical representations that revealed the
intricate behavior of the system. Furthermore, the program delved into the
investigation of transmission resonances, offering the exact energy of the
particle associated with the total transmission. Furthermore, by modifying
the code, the program explored the impact of impurities in the scattering
process. This research and the associated program contribute to our
understanding of quantum scattering and provide a valuable tool for studying
the behavior of quantum systems involving multiple Dirac $\delta -$
potentials in a wide range of scenarios. In conclusion, our paper represents
a significant contribution to the study of scattering problems in quantum
mechanics, particularly in the context of impurities in one-dimensional
systems with multiple Dirac $\delta -$potentials. The development of the
Python-based user interface program further enhances the accessibility and
accuracy of analyzing and visualizing such systems. As we progress in our
understanding of quantum mechanics, the knowledge gained from this research
paves the way for future advancements and applications in various scientific
and technological domains.

\textbf{Data Availability Statement} \newline
Data sharing does not apply to this article as no datasets were generated or
analyzed during the current study.


\begin{thebibliography}{99}
\bibitem{E1} R. de L. Kronig and W. G. Penney, \textit{A Quantum Mechanics
of Electrons in Crystal Lattices}, Proc. R. Soc. \textbf{130}, 499 (1931).

\bibitem{E2} Q. Jieli, R. Zheng, and L. Zhou. \textit{Bound states of
spin-orbit coupled cold atoms in a Dirac delta-function potential,} Journal
of Physics B: Atomic, Molecular and Optical Physics \textbf{5, }12 (2020).

\bibitem{E3} C. J. Pethick and H. Smith, \textit{Bose-Einstein Condensation
in Dilute Gases Cambridge University Press,} Cambridge, (2008).

\bibitem{D1} F. Erman and S. Seymen, \textit{A direct method for the low
energy scattering solution of delta shell potentials}, Eur. Phys. J. Plus 
\textbf{137}, 308 (2022).

\bibitem{D2} A. H. Ardila and T. Inui, \textit{Threshold scattering for the
focusing NLS with a repulsive Dirac delta potential}, Journal of
Differential Equations \textbf{313,} 54 (2022).

\bibitem{D3} S. Jarosz and J.Vaz Jr., \textit{Bound and scattering states
for supersingular potentials,} Annals of Physics \textbf{434}, 168617 (2021).

\bibitem{D4} S. Mudra and A. Chakraborty, \textit{Propagator calculations
for time dependent Dirac delta potentials and corresponding two state models,%
} Physics Letters A \textbf{418}, 127725 (2021).

\bibitem{D5} S. Fassari, M. Gadella, L. M. Nieto, et al. \textit{The Schr%
\"{o}dinger particle on the half-line with an attractive }$\delta $\textit{%
-interaction: bound states and resonances,} Eur. Phys. J. Plus \textbf{136},
673 (2021).

\bibitem{D6} B. Cheng, Y.-M. Chen, C.-F. Xu, D.-L. Li, and X.-G. Deng, 
\textit{Nonlinear Schr\"{o}dinger equation with a Dirac delta potential:
finite difference method}, Commun. Theor. Phys. \textbf{72}, 025001 (2020).

\bibitem{D7} T. Sandev, I. Petreska and E. K. Lenzi, \textit{Constrained
quantum motion in }$\delta $\textit{-potential and application of a
generalized integral operator,} Computers and Mathematics with Applications 
\textbf{78}, 1695 (2019).

\bibitem{D8} J. M. M.-Casta\~{n}eda, L. M. Nieto and C. Romaniega, \textit{%
Hyperspherical }$\delta -\delta ^{\prime }$\textit{\ potentials,} Annals of
Physics \textbf{400}, 246 (2019).

\bibitem{D9} F. Erman, M. Gadella, and H. Uncu, \textit{On scattering from
the one-dimensional multiple Dirac delta potentials,} Eur. J. Phys. \textbf{%
39}, 035403 (2018). 

\bibitem{E4} M. Belloni and R. W. Robinett, \textit{The infinite well and
Dirac delta function potentials as pedagogical, mathematical and physical
models in quantum mechanics,} Phys. Rep. \textbf{540}, 25 (2014).

\bibitem{E5} Y. N. Demkov and V. N. Ostrovskii, \textit{Zero-range
Potentials and Their Applications in Atomic Physics,} Plenum Press, New
York, (1988).

\bibitem{E6} M. Belloni, and W. R. Richard, \textit{The infinite well and
Dirac delta function potentials as pedagogical, mathematical and physical
models in quantum mechanics,} Phys. Rep. \textbf{540}, 2 (2014)

\bibitem{E7} V. E. Barlette, M. M. Leite, S. K. Adhikari, \textit{Integral
equations of scattering in one dimension,} Am. J. Phys. \textbf{69}, 1010
(2001).

\bibitem{E8} D. Lessie and J. Spadaro, \textit{One dimensional multiple
scattering in quantum mechanics,} Am. J. Phys. \textbf{54}, 909 (1986).

\bibitem{E9} I. R. Lapidus, \textit{Resonance scattering from a double
delta-function potential,} Am. J. Phys. \textbf{50}, 663 (1982).

\bibitem{E10} P. Senn, \textit{Threshold anomalies in one dimensional
scattering,} Am. J. Phys. \textbf{56}, 916 (1988).

\bibitem{E11} P. R. Berman, \textit{Transmission resonances and Bloch states
for a periodic array of delta function potentials,} Am. J. Phys. \textbf{81}%
, 190 (2013).

\bibitem{E12} G. Cordourier-Maruri, R. De Coss, and V. Gupta, \textit{%
Transmission Properties of the one-dimensional array of delta potentials,}
Int. J. Mod. Phys. B \textbf{25}, 1349 (2011).

\bibitem{E13} Z. Ahmed, S. Kumar, M. Sharma, V. Sharma, \textit{Revisiting
double Dirac }$\delta -$\textit{potential,} Eur. J. Phys. \textbf{37},
045406 (2016).

\bibitem{E14} S. H. Patil, Quadrupolar, \textit{triple delta-function
potential in one dimension,} Eur. J. Phys. \textbf{30}, 629 (2009).

\bibitem{E15} P. Pereyra, and C. Edith, \textit{Theory of finite periodic
systems: general expressions and various simple and illustrative examples,}
Phys. Rev. B \textbf{6, }.20 (2002)

\bibitem{E16} P. Pereyra, \textit{The transfer matrix method and the theory
of finite periodic systems. From heterostructures to superlattices,} Basic
Solid State Physics B, \textbf{259}, 3 (2022)

\bibitem{E17} P. Markos, and C. M. Soukoulis,\textit{Wave propagation: from
electrons to photonic crystals and left-handed materials,} Princeton
University Press, 2008.

\bibitem{E18} R. P-Alvarez, and H. R-Coppola. \textit{Transfer Matrix in 1D
Schr\"{o}dinger Problems with Constant and Position-Dependent Mass,} Basic
Solid State Physics B, \textbf{145}, 2 (1988).

\bibitem{E19} M. I.-Reyes, R. P. Alvarez, and I. R-Vargas, \textit{Transfer
matrix in 1D Dirac-like problems,} Journal of Physics: Condensed Matter 
\textbf{35}, 39 (2023).

\bibitem{E20} D. W. L. Sprung, G. V. Morozov, and J. Martorell. \textit{%
Antireflection coatings from analogy between electron scattering and spin
precession,} J. App. Phys. \textbf{93}, 8 (2003)

\bibitem{E21} M. Coquelin, C. Pacher, M. Kast, G. Strasser, and E. Gornik, 
\textit{Wannier-Stark level anticrossing in biperiodic superlattices,} Basic
Solid State Physics B, \textbf{243}, 14 (2006).

\bibitem{E22} D. W. L. Sprung, L. W. A. Vanderspek, W. van Dijk, J.
Martorell, and C. Pacher, \textit{Biperiodic superlattices and the
transparent state,} Phys. Rev. B, \textbf{77}, 035333 (2008).

\bibitem{GitHub} \textbf{%
\url{https://github.com/Erfan-keshavarz/QuantumScatteringDiracPotential}.}
\end{thebibliography}
\end{document}